\UseRawInputEncoding
\documentclass[a4paper,10pt]{article}

\usepackage{graphicx}

\usepackage{bm,dsfont}
\usepackage{amsmath}

\def\ti#1{\tilde{#1}{}}\def\wti#1{\widetilde{#1}{}}

\def\bma{{\bm a}}\def\bmk{{\bm k}}

\def\bmC{{\bm C}}
\def\bmD{{\bm D}}\def\bmF{{\bm F}}\def\bmG{{\bm G}}
\def\bmK{{\bm K}}\def\bmP{{\bm P}}
\def\bmT{{\bm T}}

\def\ga{\alpha}\def\gb{\beta}\def\gve{\varepsilon}\def\gk{\kappa}
\def\gs{\sigma}\def\gtt{\theta}
\def\gD{\Delta}

\def\re{\mathrm{e}}\def\ri{\mathrm{i}}

\def\one{\mathds{1}}\def\RR{\mathds{R}}\def\CC{\mathds{C}}

\def\kost#1#2{(~\!#1~\!|\!~#2~\!)}

\def\abs#1{\vert#1\vert}
\def\d={\buildrel \rm def \over =}

\def\kost#1#2{(~\!#1~\!|\!~#2~\!)}

\def\smatD#1#2{\left(\begin{smallmatrix}#1\\ #2\end{smallmatrix}\right)}

\begin{document}

\title{Irreducible and site symmetry induced representations of single/double ordinary/gray layer groups}

\author{B. Nikoli\'c$^\mathrm{a}$, I. Milo\v sevi\'c$^\mathrm{a}$\footnote{Corresponding author: ivag@rcub.bg.ac.rs}, T. Vukovi\'c$^\mathrm{a}$, N. Lazi\'c$^\mathrm{a}$, S. Dmitrovi\'c$^\mathrm{a}$,\\ Z. Popovi\'c$^\mathrm{a}$ and M. Damnjanovi\'c$^\mathrm{a,b}$\\ \phantom{a} \\
$^\mathrm{a}$ NanoLab, Faculty of Physics, University of Belgrade,\\ Studentski trg 12, Belgrade, Serbia and \\
$^\mathrm{b}$ Serbian Academy of Sciences and Arts,\\ Kneza Mihaila St. 35, Belgrade, Serbia}

\maketitle

\begin{abstract}
Considered are eighty sets of layer groups, each set consisted of four groups: ordinary single and double, and gray single and double layer group. Structural properties of layer groups (factorization onto cyclic subgroups and existence of grading according to the sequence of halving subgroups) enable efficient symbolic computation (by POLSym code) of the relevant properties, real and complex irreducible and allowed (half-)integer (co-)representations in particular. This task includes, as the first step, classification of the irreducible domains based on the group action in Brillouin zone combined with torus topology. Also, the band (co-)representations induced from  the irreducible (co-)representations of Wyckoff position stabilizers (site symmetry groups) are decomposed onto the irreducible components. These, and other layer group symmetry related theoretical data relevant for physics, layered materials in particular,  are tabulated and made available through the web site.
\end{abstract}

\section{Introduction}

Over last decade interest in layered structures has increased tremendously. Graphene and related structures and then flatlands beyond graphene have been discovered and systematically studied. Layered materials have been thoroughly investigated and their fundamentally and technologically important properties have been revealed. Recent research breakthrough relating topological phases with symmetry, through the graph analysis of the band representations give strong momentum to symmetry based research in solid state physics. While space groups,   symmetries of 3D crystals, are established knowledge,  symmetry groups of layered structures are not so exhaustively  elaborated. Namely, first layer groups (LGs) classification, their generators and Wyckoff positions are published~\cite{EWood} while only forty years later, in International Tables for Crystallography~\cite{KLE}, tabulated are myriad of the relevant data including various sets of generators, positions (Wyckoff symbol, multiplicity, site symmetry, coordinates), asymmetric units, group generators, maximal subgroups and minimal supergroups and other direct space related data. However, reciprocal space and irreducible representations (IRs) are rarely discussed. In fact, for each LG there are supergroups among the space groups, and data for the supergroup (irreducible domains, IRs, etc.) can  in principle be used~\cite{EVARESTOVSiteSym} to derive the corresponding data for the LG~\cite{LSiteSym19}. Being cumbersome, this procedure is not straightforward and therefore,  IRs of LGs have been derived independently~\cite{IDGIR}. More than a decade ago  Bilbao crystallographic server~\cite{DATABilbao,PSBilbao} devoted a section to the LGs, which has recently been amended by the band representations, being induced from the representations of the site symmetry groups~\cite{LSiteSym19}. However, a bunch of information is still missing, and the present paper is aimed at completing the field.

Therefore,  we generate exhaustive data set, covering the following types of LGs: single and double groups (relevant for spinless and spinful systems), as well as ordinary and gray groups (systems without and with time reversal symmetry, TRS). For all of these groups their IRs (for gray groups irreducible co-representations, co-IRs), and band (co-)represenations are derived. It should be stressed out that all derivations are performed by computer algebra, meaning that a number of subroutines are developed (induction for general points of Brillouin zone, reduction of various types of representations over IRs parameterized by wave vector, including construction of band representations, and their reduction) and gathered within POLSym code~\cite{YILG}. This means that beside the presented database, which are already directly accessible,  there are additional data (black-and-white LGs data, e.g.) which at present can be obtained on request only (as the online platform is currently under construction).

\section{Layer groups: structure and generators}\label{SGens}
Gathering symmetries of quasi-2D crystals, LGs are generated by translations (conventionally spanning $xy$-plane) and additional orthogonal transformations. In a wider sense, when half-integer spin space is also considered, orthogonal elements have twofold covering by ${\bm SU}(2)$ matrices, forming double groups altogether. Analogously, these (so called ordinary single and double groups) are extended into the magnetic ones, when time reversal is included.

\subsection{Single ordinary layer groups}\label{SSingle}
Single ordinary LGs are  diperiodic subgroups of the Euclidean group $E(3)=\bmT\wedge {\bm O}(3)$ (semidirect product). Since the lattice is invariant under the groups' transformations, each LG $\bmG$ has invariant translational subgroup $\bmT=\bmT(\bma_1,\bma_2)$, where the lattice periods (basis) $\{\bma_1,\bma_2\}$ span $xy$-plane.
General element is (Koster--Seitz notation) $g=\kost{A}{\bma}$, where $A$ is three-dimensional orthogonal transformation preserving $xy$-plane (and $z$-axis) and $\bma$ is a translational vector (belonging to the $xy$-plane); the transformations $\kost{A}{0}$ will be written as $A$. The introduced setup implies that $A$ can be represented (in a basis of the two vectors lying in the $xy$-plane, and the third one along $z$-axis) by a block-diagonal ${\bm O}(3)$ matrix, with the upper $2\times2$ block  from ${\bm O}(2)$, while $A_{33}=\pm1$. Then, all transformations have linear representation by $4\times 4$ matrices of the block form $E(g)=E\kost{A}{\bma}=\smatD{A&\bma}{\vec{0}&1}$; $\bma$ is 3D column of the translational vector components, while $\vec{0}=(0,0,0)$. The representation is faithful, and therefore the transformations are naturally identified with these matrices.

The key structural feature of single LGs (either ordinary or magnetic) is their factorization into  weak direct  product of cyclic groups~\cite{IDGIR}:
\begin{equation}\label{EGWDP}
\bmG=\bmC(g_1)\circ\cdots\circ\bmC(g_{[\bmG]}),
\end{equation}
where $\bmC(g_i)$ is a cyclic group generated by $g_i$, and $[\bmG]$ is the number of cycles. The elements $\{g_1,\cdots,g_{[\bmG]}\}$ will be called \emph{factorization generators} (the choice of the cycles in \eqref{EGWDP}, may not be unique; in addition, the generators $g_i$ are  not unique for cycles of the length greater than 2).
A useful consequence of such a structure is that each element is expressed uniquely as a monomial over generators:
\begin{equation}\label{Egmono}
g=g^{\ga_1}_1\cdots g^{\ga_{[\bmG]}}_{[\bmG]},\quad \ga_1,\ga_2=0,\pm1,\dots,\quad\ga_3,\ga_4,\dots=0,\dots,\abs{g_i}-1,
\end{equation}
providing thus the identification of an element with its monomial exponents: $g\equiv(\ga_1,\dots,\ga_{[\bmG]})$.
The  convention here introduced should be noted: the first two generators are of infinite order (translations, screw axes or glide planes), and thus their exponents can take any integer value; otherwise, the exponents are positive integers less than the order of an element $g_i$:  $\abs{g_i}=\abs{\bmC(g_i)}$.

\emph{Graded generators} are another set, useful in various structural analyses and construction of IRs (Subsection~\ref{SSIRs}). Namely, generators of single ordinary LGs can be arranged into four grades. First grade groups, $\bmG^{(1)}$, can be factorized as a semi-direct product, $\bmT\wedge\bmC_n$, with pure translational generators $\ell_1=\kost{\one}{\bma_1}$ and $\ell_2=\kost{\one}{\bma_2}$, and rotation $\ell_3=C_n$, as the third generator. Thus, the first grade groups are  LG1, LG3, LG65, LG49 and LG73,
corresponding to the orders $n=1,2,3,4,6$ of the principle axis of rotation (note that $\ell_3=C_1$ for $n=1$). Each LG $\bmG$ has one maximal subgroup $\bmG^{(1)}$ of the first grade. Second grade groups can be built expanding the first grade ones: $\bmG^{(2)}=\bmG^{(1)}+\ell_4\bmG^{(1)}$ by adding a generator, $\ell_4$, with the property $\ell^2_4\in\bmG^{(1)}$. Further expansions (by a generator,  the square of which belongs to the second grade) can be continued until all eighty LGs are constructed (single ordinary groups). In this way  series $\bmG^{(i)}$ of subgroups is formed for each LG $\bmG$, where the last in the series is $\bmG=\bmG^{(\bmG)}$ (number
$(\bmG)$ will be called \emph{grade of the group}), and $i$-th group of the series, $\bmG^{(i)}$, halves the next one, $\bmG^{(i+1)}$. In general, the graded generators are not the minimal set of generators; thus instead of \eqref{Egmono}, the  unique expansion over graded generators
\begin{equation}\label{ElmonoGrad}
 g=\ell_1^{\gb_1}\dots\ell_{(\bmG)+2}^{\gb_{(\bmG)+2}},\ \
\gb_1,\gb_2=0,\pm1,\dots;\  \gb_3=0,\dots,n-1;\
\gb_4,\dots,\gb_{(\bmG)+2}=0,1
\end{equation}
should be used.

Each generator carries one or two quantum numbers: translations are associated to momentum, $\bmk$, rotation $C_n$ to the $z$-component of the angular momentum $m$, while the involutions $U$ (rotation for $\pi$ around horizontal axis), $\gs_v$ and $\gs_h$ (vertical and horizontal mirror planes) give the parity quantum numbers $\Pi_U$, $\Pi_v$ and $\Pi_h$ (sometimes shortened to $U$, $V$ and $H$); in addition, the graded generators with $i>3$ (with allowed exponents $0,1$) which are not involutions carry pairs of quantum numbers: $k$ and $\Pi_U$ for helical axes, $k$ and $\Pi_h$ or $\Pi_v$ for (horizontal and vertical) glide planes, and $m$ and $h$ for $S_{2n}=C_{2n}\gs_h$. Moreover, in the last case the square $\ell^2_i=\ell'$ is one of the precedent graded generators. If $\ell'$ is substituted by  $\ell_i$ (allowing for $\gb_i$ all values from the cycle of $\ell_i$), the complete  group will be generated again; actually, the factorization
generators used (at the web site and in the POLSym code) are obtained by such reduction of the set of graded generators. For instance, graded set of generators of a nonsymmorphic LG7 are $\{\ell_1=\kost{\one}{\bma_1},\ell_2=\kost{\one}{\bma_2},\ell_3=C_2,\ell_4=\kost{\gs_h}{\bma_2/2}\}$. The first three elements generate first grade (halving) subgroup LG3, while the whole group LG7 is generated by $\{g_1=\ell_1,g_2=\ell_4,g_3=\ell_3\}$.

\subsection{Clusters of layer groups}\label{SSBundle}

As it is well known, inclusion of a half-integer spin degree of freedom urges for  rotational group ${\bm SO}(3)$ extension  to its universally covering group ${\bm SU}(2)$, which is further  inherited by all groups containing rotations. These double coverings are known as double groups. Also, when TRS is considered, besides the ordinary groups (single and double) gray and black-and-white magnetic groups can be associated to the same LG. Thus the  eighty sets of LGs include ordinary LG $\bmG$, and the corresponding double LG $\wti{\bmG}$, gray LG $\bmG1'$, double gray LG $\wti{\bmG}1'$, magnetic and double magnetic groups making the \emph{clusters} associated to each of the eighty ordinary single LGs. Since 368 black-and-white groups are not analyzed here, for the time being presented are only the four-membered ($\bmG$, $\wti{\bmG}$, $\bmG1'$, $\wti{\bmG}1'$) parts of each of the clusters.

In 2D spin-half space translations and spatial inversion act trivially and thus only orthogonal transformations' part $R=\det(A)A$ of $g=\kost{A}{\bma}$ is  relevant. However, since  ${\bm SU}(2)$  is a double covering group of the ${\bm SO}(3)$ rotational group, two antipodal matrices $u(g)$ and $-u(g)$ are associated to each Euclidean transformation $g$. Double LG $\wti{\bmG}$ is double extension of the single LG $\bmG$, so that its action is expanded beyond the Euclidean  action in $\RR^3$: it acts also in the  spin space  $\CC^2$. Therefore, two \emph{antipodal} elements $\ti{g}=(g,u(g))$ and $\ti{g}'=(g,-u(g))$ are associated to each element $g\in\bmG$. Note that either of the two opposite-sign matrices, mapped by two-to-one homomorphism of ${\bm SU}(2)$ onto ${\bm SO}(3)$ into $R$, can be taken as $u(g)$. Consequently, elements of double group are faithfully represented by 6D matrices $\ti{E}(\ti{g})=E\kost{A}{\bma}\oplus u(A)$~\cite{DLG}. For double groups pairwise intersections of cycles in \eqref{EGWDP} may contain besides identity $\ti{e}$, also its antipodal  $\ti{e}'$, hence this is not a weak direct product, which prevents uniqueness of the exponents in \eqref{Egmono}. Fortunately, this can be fixed by adding antipodal to identity $\ti{e}'$ of order two as the last generator~\cite{DLG}:
\begin{equation}\label{Egmono2}
\ti{g}=\ti{g}^{\ga_1}_1\ti{g}^{\ga_2}_2\cdots \ti{g}^{\ga_{[\bmG]}}_{[\bmG]} \ti{e}^{\prime\gve                    },\quad \ga_1,\ga_2=0,\pm1,\dots,\ \ \ga_3,\ga_4,\dots=0,\dots,\abs{g_i}-1,\gve=0,1.
\end{equation}

Finally, in magnetic (single or double) group $\bmG t'=\bmG+t\gtt\bmG$ there is additional generator $t\gtt$ such that $(t\gtt)^2=t^2$ is an element of $\bmG$. Here we consider only the case of gray groups, when $t$ is the identity; otherwise, black-and-white groups are obtained. Clearly, this introduces additional generator $\gtt$.

Note that in gray groups $\gtt$ is placed as  the last generator (by convention), while $\ti{e}'$ (double groups) is put right after the geometrical transformation generators.

\section{Brillouin zone and irreducible domain}\label{SBZIDs}

Irreducible representations of group $\bmT$ are classified by quasi-momentum $\bmk$ running over Brillouin zone (BZ) which has a form of a torus $T^2$.  The group is abelian, and IRs are 1D: $\gD^{(\bmk)}\kost{\one}{\bma_i}=\re^{\ri\bmk_i\bma_i}$ ($i=1,2$).
They are starting point for construction of IRs of a LG $\bmG$. In the induction procedure, the first step is to determine orbits of conjugation of $\gD^{(\bmk)}(\bmT)$ by cosets of $\bmT$, which effectively reduces to the $\bmk^*$-stars (orbits) of the polar vector action of $\bmG$ in BZ. Set of star-representatives forms an irreducible domain (ID). The latter consists of the generic (dense) stratum (with minimal stabilizer $\bmG_\bmk$),  which can be fully or partially bounded. The special strata (having stabilizers which are nontrivial supergroups of the generic stratum stabilizer $\bmG_\bmk$) are on the boundary of the generic one. Here, two significant characteristics should be highlighted. Firstly, as translations in BZ act trivially, they  are subgroups of  the all stabilizers, and thus the stabilizers are LGs themselves. Consequently, only isogonal point groups $\bmP_I$ are relevant, and thus  stabilizing elements pertaining to $\bmP_I$  are to be extended by the translations to get the full little group. Secondly, isogonal transformations effectively act as $2\times2$ upper block of the usual polar ${\bm O}(3)$ action, i.e. they form a subgroup of ${\bm O}(2)={\bm SO}(2)+U\,{\bm SO}(2)$ matrices, and  transformations differing by action in $z$-direction only are mutually identical.  Hence, there remain ten algebraically different point groups, which by stratification of BZ form IDs: $\bmC_1$, $\bmC_2$, $\bmC_3$, $\bmC_4$, $\bmC_6$ and $\bmD_1$, $\bmD_2$, $\bmD_3$, $\bmD_4$, $\bmD_6$.

\begin{figure}[]
\includegraphics[width=0.98\textwidth]{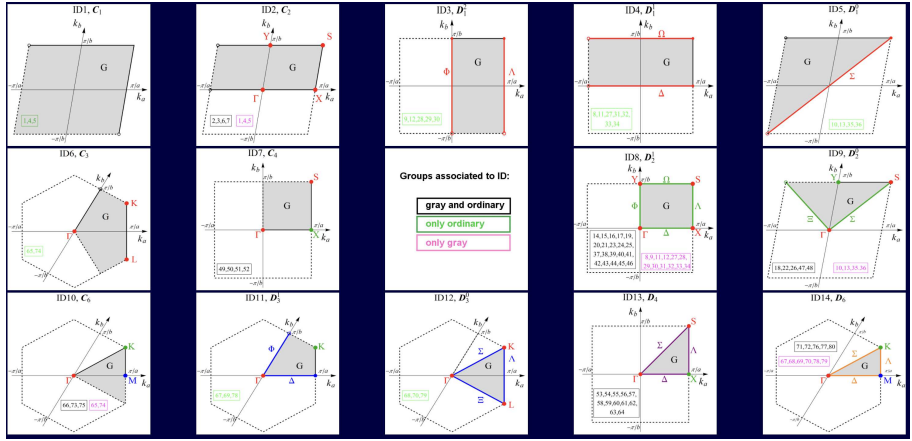}
\caption{Irreducible domains of layer groups. Gray - generic stratum (G), with black boundaries belonging to G.
Color denotes stratum relative order $\abs{\bmG}/\abs{\bmG_K}$: 1 (red) ,2 (green), 3 (blue), 4 (violet) and 6 (orange). Groups corresponding to IDs are indicated by colors (explained in the central panel).}
\label{FIDs}
\end{figure}

Further, a nontrivial topological characteristic related to the action of the two-fold horizontal rotations $U\in\bmD_{n}$ on BZ-torus should be pointed out. Namely, the $U$-axis can be  either along $\bmk_1$ or $\bmk_2$  or bisector $\bmk_0=\bmk_1+\bmk_2$ and these $U_i$-axes ($i=0,1,2$) are closed paths on torus. However, the axes are not homotopy equivalent. Namely, as  $\bmD_n$ has $n$ different $U$-axes, in the cases $n=2$ and $n=3$, existence of the $U_1$-axis entails also presence of the  $U_2$-axis (i.e these two axes are equivalent),  while in the cases $n=4$ and $n=6$, all the U-axes are equivalent. To summarize, there are altogether nine topologically (and algebraically) different dihedral groups: $\bmD^1_1$, $\bmD^2_1$, $\bmD^0_1$, $\bmD^1_2$, $\bmD^0_2$, $\bmD^1_3$, $\bmD^0_3$, (the superscripts indicate the above described position of the $U$-axis within BZ) $\bmD_4$, and $\bmD_6$;  these groups, together with aforementioned $\bmC_n$ groups, make fourteen  different isogonal group actions on torus, $\bmP^T_I$, and thus BZ-stratification results in fourteen topologically different IDs, Fig.~\ref{FIDs}.
Hence, having all the ID types of LGs at disposal, one can easily associate  LGs to IDs, with  $\bmP^T_I$ being ${\bm O}(2)$-part of the isogonal group $\bmP_I$.

Time reversal transformation $\gtt$ reverses $\bmk$ vectors, which is equivalent to $\pi$-rotation around $z$-axis. Hence, inclusion of TRS manifests as the inclusion of $C_2$ into $\bmP^T_I$, enlarging $\bmC_1$ to $\bmC_2$, $\bmC_3$ to $\bmC_6$, both $\bmD^1_1$ and $\bmD^2_1$ to $\bmD^1_2$, $\bmD^0_1$ to $\bmD^0_2$ and both $\bmD^1_3$ and $\bmD^0_3$ to $\bmD_6$, while the remaining groups stay the same (as $C_2$ is their element). In this way seven ID-types of gray groups $\bmG1'$ are identified. Black-and-white groups are of the form $\bmG\ell'=\bmG+\ell\gtt\bmG$, where $\ell$ is a Euclidean transformation preserving $xy$-plane, such that $\ell^2\in\bmG$. For these groups ID is found in an analogous way as for the gray groups (for which $\ell=\one$, or equivalently $\ell\in\bmG$): if $\bmP^T_I$ corresponds to $\bmG$, then addition of $C_2\ell$ to $\bmP^T_I$ gives the toroidal isogonal group of a black-and-white group.

Clearly, $\bmG$ (ordinary or magnetic) and its double group $\wti{\bmG}$ have the same ID, as the extension by spin space does not affect the translational group (and its IRs parameterized by the BZ wave vector).

In  conclusion,  the LGs action yield fourteen ID types, each ID decomposes onto strata, counted by $K$.  Only generic stratum is 2D and dense. Lower-D strata, special lines and special points make (partial) boundary of the generic one. The sets of LGs associated to the specific IDs are listed in Table~\ref{TIDsGr}.

\begin{table}
\caption{Irreducible domains of layer groups. Columns: counter (ID), topological class of the 2D isogonal group ( $\bmP^T_I$), ordinals of associated ordinary single and double groups ($\bmG$, $\wti{\bmG}$), IDs of the corresponding gray groups (ID of $\bmG1'$ {\&}  $\wti{\bmG}1'$), bolded when different from the IDs of the counterparts (without TRS, ID-column).}
\label{TIDsGr}
\begin{tabular}{lllc}      % Alignment for each cell: l=left, c=center, r=right
ID&$\bmP^T_I$&$\bmG$, $\wti{\bmG}$&ID of $\bmG1'$ {\&}  $\wti{\bmG}1'$  \\ \hline
1 &$\bmC_1$  &1,4,5                         &\textbf{2}\\
2 &$\bmC_2$  &2,3,6,7                       &2    \\
3 &$\bmD^2_1$&9,12,28,29,30                 &\textbf{8}\\
4 &$\bmD^1_1$&8,11,27,31,32,33,34           &\textbf{8}\\
5 &$\bmD^0_1$&10,13,35,36                   &\textbf{9}\\
6 &$\bmC_3$  &65,74                         &\textbf{10}\\
7 &$\bmC_4$  &49,50,51,52                   &7           \\
8 &$\bmD^1_2$&14,15,16,17,19,20,21,23,24,25,&8       \\
  &          &37,38,39,40,41,42,43,44,45,46 &            \\
9 &$\bmD^0_2$&18,22,26,47,48                &9         \\
10&$\bmC_6$  &66,73,75                      &10        \\
11&$\bmD^1_3$&67,69,78                      &\textbf{14}\\
12&$\bmD^0_3$&68,70,79                      &\textbf{14}\\
13&$\bmD_4$  &53,54,55,56,57,58,            &13          \\
  &          &59,60,61,62,63,64             &            \\
14&$\bmD_6$  &71,72,76,77,80                &14    \\
\end{tabular}\end{table}

\section{Representations}\label{SReps}

Group representations are well defined on the generators only, as homomorphism of the representation and unique monomial expansion \eqref{Egmono}  give:
\begin{equation}\label{EDmono}
D(g)=D^{\ga_1}(g_1)D^{\ga_2}(g_2)\cdots D^{\ga_{[\bmG]}}(g_{[\bmG]}),
\end{equation}
for  arbitrary group element $g$.
Here, in addition to the above introduced Euclidean representation $E\kost{A}{\bma }$ and natural spin representation $u({\ti g})$, considered are IRs and allowed representations. The latter are representations of stabilizers, which are also LGs, i.e groups of infinite order. However, since pure translations are represented as (multiple of) $\gD^{(\bmk)}$,  to avoid non-unique choice of generators, they are given on the elements of the little co-group (which forms transverzal of $\bmT$ in the little group).

\subsection{Irreducible representations}\label{SSIRs}

Standard prescription for construction of IRs uses ID in a way that for each representative $\bmk$  of a connected stratum $K$ its little group (stabilizer) $\bmG_K$ is found.  Then the allowed representations (by definition, IRs of $\bmG_K$ subducing on $\bmT$ multiple of $\gD^{(\bmk)}(\bmT)$) are constructed.  As the little group is a LG itself, the little co-group $\bmG_K/\bmT$ is a crystallographic axial point group, and its projective IR  $d^{(K\gk)}(\bmG_K/\bmT)$ is one factor of the allowed representation, while the other one is a phase related to $\gD^{(\bmk)}(\bmT)$ (compensating the factor system of the aforementioned projective representation). Finally,  each of these representations $d^{(K\bmk\gk)}(\bmG_\bmK)$ gives by induction~\cite{MACKEYIndI,WIG59,TJANSSEN} one IR of the whole group: $D^{(K\bmk\gk)}(\bmG)=d^{(K\bmk\gk)}(\bmG_K)\uparrow\bmG$; gathering all of them for each $K$ in ID, the complete set of non-equivalent IRs of $\bmG$ is obtained. IRs are grouped in the $K\gk$-series of the representations of the same form, differing in the value of $\bmk$, with the same (projective) representation $d^{(K\gk)}(\bmG_K/\bmT)$.

An alternative procedure encoded in POLSym utilizes graded generators, and skips the allowed representations, except for the simplest, first grade LGs $\bmG^{(1)}$. These groups are symmorphic, with little co-groups $\bmC_{n_K}$, where $n_K$ divides $n$ (for generic stratum $n_K=1$, while the remaining strata are special points for which $n_K>1$), having only one-dimensional linear (trivial projective) IRs.  Hence, allowed IRs of the stabilizers are $d^{(K\bmk\gk)}(\ell_{i})=\gD^{(\bmk)}(\ell_{i})$ ($i=1,2$) and  $d^{(K\bmk\gk)}(\ell_3=C_n)=\re^{\ri2\gk\pi/n_K}$, where $\gk$ is integer or (for double groups) half-integer within the interval $(-n_K/2,n_K/2]$. The induction procedure gives $n/n_K$-dimensional matrices. For arbitrary group $\bmG$ IRs are obtained from IRs of its (unique) first grade subgroup $\bmG^{(1)}$. Namely, as $\bmG^{(1)}$ is a halving subgroup in  $\bmG^{(2)}$, IRs of the latter are straightforwardly found from IRs of $\bmG^{(1)}$ using induction procedure for index-two supergroups~\cite{JANSEN-BOON,YILG}. The same method is successively applied in the follow-up grade(s), until  IRs of $\bmG$ are determined. For magnetic groups, the final step is the so called $*$-induction~\cite{JANSEN-BOON,YILG} of co-IRs of a magnetic group from IRs of its halving subgroup. It is important  to find the strata of $\bmG$ first and then in each step IRs of $\bmG^{(i)}$ for each of the  strata of $\bmG$  are constructed. The additional symmetries of the higher-grade groups may refine ID, in a way that  special lines and special points of the preceding grade group are retained, while the new ones may emerge.

\subsection{Allowed representations}\label{SARs}

Allowed representations are of relevance for many physical applications and since in the here applied induction-procedure their construction is skipped, the method of extracting allowed representations from IRs is developed.  At first, for the representations associated to the $K$-stratum, set of IRs $D^{(K\bmk\gk)}(\bmG)$ is selected. Each element $f$ of the stabilizer $\bmG_K$ can be expressed in terms of the group generators in the form \eqref{Egmono}, and thus the matrices $D^{(K\bmk\gk)}(f)$ are easily found using \eqref{EDmono}. As has been previously emphasised, little group is a LG and thus there are infinitely many such elements. However, the stabilizers are characterized by the stabilizer co-group elements, i.e.  the coset representatives of the translational subgroup $\bmT$ in $\bmG_K$. The orbit order (star $\bmk^*$ of $\bmk$ from $K$) is  $\abs{\bmk^*}= \abs{\bmG/\bmG_K}$ (equal to the ratio of the isogonal groups of $\bmG$ and $\bmG_K$), and the above described induction-procedure from the little group allowed representation enlightens that when $D^{(K\bmk\gk)}(f)$ is partitioned into $\abs{D}/\abs{\bmk^*}=\abs{D}/\abs{\bmG/\bmG_K}$-dimensional blocks (here, $\abs{D}$ denotes the dimension of the representation $D$ and $\abs{d^{(K\bmk\gk)}}= \abs{D}/\abs{\bmG/\bmG_K}$ is order of the coset space $\bmG/\bmG_K$), the first block gives the allowed representation: $d^{(K\bmk\gk)}(f)=D^{(K\bmk\gk)}_{11}(f)$.

\section{The web site}\label{SSite}

All here described results (with a number of the relevant extensions) are presented at the web-site www.nanolab.group/layer (except for the clusters 58 and 80 a free registration is needed).

\begin{figure}[]
\includegraphics[width=0.98\textwidth]{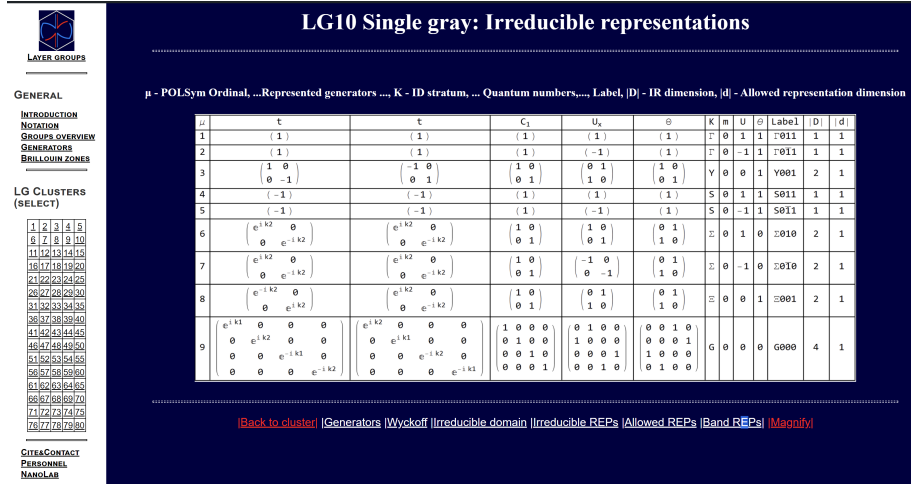}
\caption{Illustration of the web site information: co-IRs of single gray LG10 and  RIRs (if the $\theta$-columns are disregarded) of single ordinary LG10.}
\label{FIRs}
\end{figure}

The home page \framebox{Layer groups} contains  menu (Fig.~\ref{FIRs}, left margin), with \framebox{Introduction} gathering several notes on the purpose and main ideas of the site, \framebox{Notation} giving the system of labeling (group transformations and Brillouin zone strata, in particular) with comparison to other referent sources, as well as definitions of the not so frequently used notions. Follows page \framebox{Groups overview} with general data on the LGs: their ordinal and international symbol~\cite{KLE}, holohedry, factorization, isogonal group, grade $(\bmG)$, i.e. the number of graded generators minus 2, ID, symmorphism, the factorization generators that are used in POLSym. Generators are given in the standard order: firstly the translational ones (including possible fractional translations), then, depending on the group,  $g_3$ is either $C_n$ or $S_{2n}=C_{2n}\gs_{\mathrm{h}}$, and the remaining generators (at most two for ordinary single groups) follow. To remind, for double groups  the next one is the antipodal to identity, $\ti{e}'$. In gray groups the last generator is time reversal $\theta$.

Generators are specified at page \framebox{Generators} where listed are all generators (22 in total) used for LGs (including double and gray ones). The first eight generators comprise lattice and fractional translations, and therefore contain lattice parameters. Analogously, generators 9 and 15, $C_n$ and $S_{2n}$ depend on the principle axis order; all these parameters are to be substituted when used in a particular group. Then  matrices of Euclidean and natural spin representations follow. Euler angles are the next entry. In fact, orthogonal part $A$ of a generator is either $R$ or $JR$; while $J$ is spatial inversion, $R$ is a rotation from ${\bm SO}(3)$ defined by three Euler angles (real numbers) complemented  with number 1 if $A$ is a rotation,  and  with $-1$, otherwise. The orders of the generators in Euclidean and sum of the Euclidean and spin space, as well as conventional order of the latter in the described notation which includes $\ti{e}'$ follow. Finally, quantum numbers introduced by the generators are given: $\bmk$ for translations, $m$ for rotations, and different parities. Glide planes and screw axes, in addition to $\bmk$, introduce one more quantum number, associated to the orthogonal element. Analogously, $S_{2n}$ has rotational part yielding quantum number $m$ and reflectional part yielding parity quantum number  $H$.

\framebox{Brillouin zones} presents IDs of LGs.

Lastly, there is a  table with links to eighty clusters  of LG \framebox{Layer group $N$}, $N=1,\dots,80$. For the time being, only data for four-membered parts of clusters is presented (ordinary single and double, and gray single and double groups, while the data on  black and white groups is under construction); these groups will be denoted here as "LG $N$ SPIN MAG" where SPIN may be single or double, and MAG ordinary or gray. For each of them, at a separate linked page, given are the following data:

\begin{description}
\item{\framebox{LG $N$ SPIN MAG: Generators}} Data on generators of a particular group: Euclidean and natural spin representation, Euler angles, quantum numbers,  order of the element used in \eqref{Egmono}, and for double groups given is also the order of the element applied in \eqref{Egmono2}.

    \item{\framebox{LG $N$ SPIN MAG: Wyckoff}} Stratification of the Euclidean space obtained by the group action, i.e. description of the fundamental domain. For each stratum its label, coordinates of the general Wyckof position, stabilizer (site-symmetry group), elements of the stabilizer (exponents of \eqref{Egmono} or \eqref{Egmono2}), order of the stabilizer. Topology: adjacency graph of strata. Maximal Wyckoff positions are singled out.

\item{\framebox{LG $N$ SPIN MAG - Brillouin zone}} Layer group action onto Brillouin zone: each stratum (generic, special lines and points) is labeled by a symbol, general $\bmk$ vector with its star, a typical numeric point of the stratum, order of little co-group and its elements (exponents of the monomial in generators), associated IRs' ordinals, with the IR dimension and allowed representations. Depicted are Brillouin zone strata and IDs.

\item{\framebox{LG $N$ SPIN MAG - IRs}} Tables of (co)-IRs. Beside ordinals (used in POLSym) the matrices ($\bmk$-dependent in general) of $K\gk$-series are given for the group generators. Also some data are presented: corresponding quantum numbers (the first one is label giving the momenta values stratum), symbol of IR based on the quantum numbers (bar denotes negative sign, underline denotes division by two) and dimension of the allowed representation.

\item{\framebox{LG $N$ SPIN MAG - Allowed representations}} Allowed representations of the little groups. For each stratum the elements of the little co-group are given (monomial exponents in generators) together with their matrix representation.

\item{\framebox{LG $N$ SPIN MAG - Site symmetry induced band representations}} Band representations induced from IRs of the site stabilizers. For each stratum its label is given together with the stabilizer (site symmetry group) $\bmF$ and the standard point group generators; then IRs $D^{(\gs)}(\bmF)$ of the stabilizer are enumerated and explicated by matrix representation of the mentioned generators, accompanied by standard quantum numbers. Finally,  frequencies of (co)-IRs of the whole group in the induced (co-)representation $D^{(\gs)}(\bmF)\uparrow\bmG$ are tabulated.

\end{description}

\section{Discussions}

Most of the calculations, including construction of IRs and allowed representations, are performed by symbolic programming with code POLSym. The key ingredients are the afore described structural properties, LG factorization and   graded hierarchy  of index-two subgroups, suitable for the application of the modified group projector technique~\cite{MGPT0}, utilized by POLSym and being also efficient for antilinear operators.

The two sets of generators are defined; one set is suited to the  factorization \eqref{EGWDP}, and the other for the graded LG series. The first two generators are translations (generalized in \eqref{EGWDP}). The third generator in \eqref{EGWDP} is $C_n$ or $S_{2n}$, while in the grade-suited convention it is  $C_n$ solely. There are many groups with $n=1$, and thus the third generator is identity. To establish universality of the notation (and subsequent derivation of the results by computer algebra), the tables at the web site (as well as the code POLSym) refer to this generator (and use it) in a standard manner, at the cost of possible redundancy, i.e. by giving a piece of a trivial information. The remaining generators are the elements which are squared into the subgroup generated by the preceding generators. Elements $\ti{e}'$ and $\theta$, when present, stand at the position $[G]+1$ and at the last position, respectively.

From  \eqref{EGWDP} it stems that each LG can be factorized onto the \emph{generalized 2D translational group} and axial point group, $\bmG=\hat{\bmT}\bmP$~\cite{IDGIR}, where the  group of generalized translations $\hat{\bmT}$ comprises the infinite part, the product $\hat{\bmT}=\bmC(g_1)\bmC(g_2)$ of 1D generalized translations leaving the $xy$-plane invariant. The latter are depicted at the page \framebox{Factorization} (the link can be found at \framebox{Groups overview} legend).

Note that the LGs obtained by  scaling of the periods (when scaling is either independent on the present orthogonal symmetries, or orthogonal symmetry imposes certain restrictions onto  the scaling factors) are isomorphic, and thus they have the same IRs. Consequently, IRs can be given in the translational period independent form, i.e. $\bmk$-components take values from the interval $(-\pi,\pi]$, while translations are dimensionless integers; alternatively, one may assume that the periods are  length units along the translational axes.

In many physical applications, especially for the sake of interpretation of the obtained physical results, real representations  are more convenient than the complex ones. Therefore, instead of complex IRs, the so called physical (or real) irreducible representations (RIRs) are used. They are of importance not only when there is time reversal invariance, but also, in the estimation of the electron-phonon coupling in layered materials~\cite{2D-EL-PH}  where it is RIRs-based analysis which explains Jahn-Teller (in)stability in 2D, as well as the absence of coupling between flexural phonons and electrons.
By definition, RIRs are representations of ordinary (single or double) group $\bmG$, with real matrix elements, which cannot be decomposed onto real representations. Technically, RIRs coincide with subduced to $\bmG$ IRs of  the gray group $\bmG1'$ and thus they can easily be read out from the given irreducible matrix representations of the gray groups, Fig.~\ref{FIRs}. Therefore they are not listed separately.

POLSym subroutines BZStabilizers and FundamentalDomain are subroutines which output the afore presented information. The relevant procedures are suitable also for black-and-white groups, and the results and related set of data are available on request, for the time being.

Finally, let us note that POLSym program has broad scope of applications, far beyond the here presented ones, and it is expected in the near future that the POLSym platform will be made available to the scientific community.

\paragraph{Synopsis}
For single and double, ordinary and gray layer groups $\bmG$, all  irreducible  (half-)integer (co-)representations $D^{(\mu)}(\bmG)$ are derived as well as the allowed (co-)representations of little groups. Band representations for all types of layer groups are induced from the irreducible representations of the site symmetry groups and decomposed onto the irreducible components. Together with many other important characteristics of the layer groups, these results are presented at the new web-site.

\end{document}